\documentclass{ws-p8-50x6-00}

\newcommand{\eqnref}[1]{Eq.~(\ref{#1})}
\newcommand{\eqnlessref}[1]{(\ref{#1})}

\def\Det{{\rm Det}}

\def\scrD{{\cal D}}
\def\lsim{\mathrel{\lower0.3em\hbox{$\stackrel{\textstyle <}{\sim}$}}}
\def\gsim{\mathrel{\lower0.3em\hbox{$\stackrel{\textstyle >}{\sim}$}}}
\def\scrK{{\cal K}}
\def\negspace{\kern -0.4em}

\def\dvec{\raise 0.3 em \hbox{$^\leftrightarrow$} \kern -0.77 em}

\def\omegahat{\hat%
	{\setbox0=\hbox{$\omega$}%
		\kern-.025em\copy0\kern-\wd0
		\kern.05em\copy0\kern-\wd0
		\kern-.025em\raise.0433em\box0}}

\begin{document}

\title{Effective String Theory of Vortices and Regge Trajectories
	of Hybrid Mesons with Zero Mass Quarks}
\author{M. Baker and R. Steinke}

\address{University of Washington, P.O. Box num, Seattle, WA zip, USA\\
	E-mail: baker@phys.washington.edu, rsteinke@u.washington.edu}

\maketitle

\abstracts{
    We show how a field theory containing classical vortex solutions
    can be expressed as an effective string theory of long distance
    QCD describing the two transverse oscillations of the
    string. We use the semiclassical expansion of this effective
    string theory about a classical rotating string solution
    to obtain Regge trajectories for mesons with zero mass quarks.
    The first semiclassical correction adds the constant $1/12$
    to the classical Regge formula for the angular momentum of
    mesons on the leading Regge trajectory. In $D$ spacetime
    dimensions, this additive constant is $(D-2)/24$.
    The excited states of the rotating string give rise to daughter
    Regge trajectories determining the spectrum of hybrid mesons.
}



\section{Introduction}

In this talk, we first express the path integral of a renormalizable
field theory containing classical vortex solutions as an effective
string theory of those vortices. The field theory itself is
an effective theory describing phenomena at distances greater
than the radius of the flux tube, whose center is the
location of the vortex.

We use the semiclassical expansion of the effective string theory
around a classical rotating string solution to find the correction
to the classical formula for Regge trajectories due to the zero
point motion of the string. We also calculate the energies of
the excited states of the rotating string to determine the
Regge trajectories of hybrid mesons with zero mass quarks.

\section{The Effective String Theory of Vortices}

The dual Abelian Higgs model, which couples dual potentials $C_\mu$ to
a complex Higgs field $\phi$, is an example of a field theory
having classical vortex solutions.~\cite{Nielsen+Olesen}
The action $S[C_\mu,\phi]$ of the theory is
\begin{equation}
S[C_\mu,\phi] = \int d^4x \left[ -\frac{1}{4} G_{\mu\nu}^2
- \frac{1}{2} \left|\left(\partial_\mu - ig C_\mu\right) \phi \right|^2
- \frac{\lambda}{4} \left( |\phi|^2 - \phi_0^2 \right)^2 \right] \,.
\label{field action}
\end{equation}
The dual coupling constant is $g=2\pi/e$, where $e$ is
the Yang--Mills coupling constant, and $G_{\mu\nu}$ is
the dual field strength tensor. The Higgs mechanism gives the
vector particle (dual gluon) a mass $M_V = g\phi_0$, and
the scalar particle a mass $M_S = \sqrt{2\lambda}\phi_0$.
The Higgs field $\phi$ vanishes on the surface $\tilde x^\mu$,
which is the location of the vortex. Electric flux is confined
to a tube of radius $a = 1/M_V$ surrounding the vortex.

This model, regarded as a dual superconductor, provides
an effective theory of long distance QCD.~\cite{Nambu,Mandelstam,tHooft}
The quark--antiquark interaction is determined by the Wilson
loop $W[\Gamma]$:
\begin{equation}
W[\Gamma] = \int \scrD C_\mu \scrD\phi \scrD\phi^* e^{iS[C_\mu, \phi]} \,.
\label{Wilson loop def}
\end{equation}
The path integral \eqnlessref{Wilson loop def} goes over all field
configurations containing a vortex sheet bounded by the loop
$\Gamma$ formed by the worldlines of the trajectories of the
quark and antiquark on the ends of the vortex.

We carry out the functional integration \eqnlessref{Wilson loop def}
in two stages:
\begin{enumerate}
\item We integrate over all field configurations
in which the vortex is located on a particular surface
$\tilde x^\mu$, where $\phi(\tilde x^\mu) = 0$. This
integration determines the action $S_{{\rm eff}}[\tilde x^\mu]$
of the effective string theory.
\item We integrate over all
vortex sheets by expressing $\tilde x^\mu$ in terms of two
physical degrees of freedom $f^1(\xi)$ and $f^2(\xi)$,
the transverse fluctuations of the rotating string,
\end{enumerate}
\begin{equation}
\tilde x^\mu(\xi) = x^\mu(f^1(\xi), f^2(\xi), \xi^1, \xi^2) \,.
\label{x param}
\end{equation}
Using the parameterization \eqnlessref{x param}, we can write the
integration  over vortex sheets as
a path integral over the surface fluctuations $f^1(\xi)$ and $f^2(\xi)$.
The path integral \eqnlessref{Wilson loop def} over
fields then takes the following form:~\cite{Baker+Steinke2,Baker+Steinke3}
\begin{equation}
W[\Gamma] = \int \scrD f^1 \scrD f^2 \Delta_{FP} e^{iS_{{\rm eff}}[\tilde x^\mu]} \,,
\label{Wilson loop eff}
\end{equation}
where
\begin{equation}
\Delta_{FP} = \Det\left[ \frac{\epsilon_{\mu\nu\alpha\beta}}{\sqrt{-g}}
\frac{\partial x^\mu}{\partial f^1} \frac{\partial x^\nu}{\partial f^2}
\frac{\partial x^\alpha}{\partial \xi^1} \frac{\partial x^\beta}{\partial \xi^2}
\right]
\end{equation}
is a Faddeev-Popov determinant, and $\sqrt{-g}$ is the square root
of the determinant of the induced metric
\begin{equation}
g_{ab} = \frac{\partial \tilde x^\mu}{\partial \xi^a}
\frac{\partial \tilde x_\mu}{\partial \xi^b} \,.
\end{equation}
The path integral \eqnlessref{Wilson loop eff} over string fluctuations
is cut off at a distance scale $a$, the radius of the flux tube.

The theory is invariant under reparameterizations $\tilde x^\mu(\xi)
\to \tilde x^\mu(\xi(\tau))$ of the vortex worldsheet. Specifying
the parameterization \eqnlessref{x param} is analogous to fixing
a gauge in a gauge theory, and has produced the Faddeev-Popov
determinant $\Delta_{FP}$ in \eqnlessref{Wilson loop eff}.
Anomalies~\cite{Polyakov:conformal_paper,Polyakov:book}
produced in string theory by evaluating integrals
over reparameterizations of the string are not present in
\eqnlessref{Wilson loop eff}, and there is no Polchinski--Strominger
term.~\cite{Pol+Strom,ACPZ} The resulting effective theory
is a two dimensional effective field theory of the
transverse fluctuations $f^a(\xi)$.

\section{The Action of the Effective String Theory}

The action $S_{{\rm eff}}[\tilde x^\mu]$ of the effective
string theory \eqnlessref{Wilson loop eff} is determined by
the integration \eqnlessref{Wilson loop def} over all field
configurations which have a vortex on the surface
$\tilde x^\mu$. All the long distance fluctuations in
$W[\Gamma]$ are contained in the integrations \eqnlessref{Wilson loop eff}
over string fluctuations, so in the leading
semiclassical approximation $S_{{\rm eff}}[\tilde x^\mu]$
is the action \eqnlessref{field action} evaluated at a solution of
the classical field equations having a vortex at $\tilde x^\mu$.

The action $S_{{\rm eff}}[\tilde x^\mu]$ can be expanded in powers of
the extrinsic curvature tensor $\scrK^A_{ab}$ of the sheet
$\tilde x^\mu$,
\begin{equation}
S_{{\rm eff}}[\tilde x^\mu] = - \int d^4x \sqrt{-g} \left[ \sigma
+ \beta \left(\scrK^A_{ab}\right)^2 + ... \right] \,.
\label{curvature expansion}
\end{equation}
The extrinsic curvature tensor is
\begin{equation}
\scrK^A_{ab} = n^A_\mu(\xi) \frac{\partial^2 \tilde x^\mu}
{\partial\xi^a \partial\xi^b} \,,
\end{equation}
where $n^A_\mu(\xi)$, $A = 1,2$ are vectors normal to
the worldsheet at the point $\tilde x^\mu(\xi)$.
The string tension $\sigma$ and the rigidity $\beta$ are
determined by the parameters of the underlying
effective field theory \eqnlessref{Wilson loop def},
whose long distance fluctuations are described by the effective
string theory \eqnlessref{Wilson loop eff}.

The extrinsic curvature is of the order of magnitude of the angular
velocity $\omega$ of the rotating string, and $\omega^2/\sigma \sim
1/J$, where $J$ is its angular momentum. The
expansion parameter in \eqnlessref{curvature expansion}
is then $1/J$. There is evidence~\cite{Baker+Ball+Zachariasen:1990,CKNPS}
that $\lambda/g^2 \simeq 1/2$,
which corresponds to a superconductor on the I-II border where
$M_V = M_S \equiv M$. For such a superconductor there is no force
between vortices.~\cite{deVega+Schaposnik} The rigidity term, which in a sense
represents the attraction or repulsion between neighboring
parts of the string, should then vanish. That is,
the rigidity $\beta$ is zero for a superconductor on the
type I-II border.~\cite{Baker+Steinke3}

The effective action \eqnlessref{curvature expansion} can therefore
be approximated by the Nambu--Goto action,
\begin{equation}
S_{{\rm eff}} = S_{{\rm NG}} \equiv -\sigma \int d^4x \sqrt{-g} \,,
\end{equation}
and the path integral for $W[\Gamma]$ becomes
\begin{equation}
W[\Gamma] = \int \scrD f^1 \scrD f^2 \Delta_{FP}
e^{i\sigma \int d^4x \sqrt{-g}} \,.
\label{Wilson Nambu Goto}
\end{equation}
In the next section we describe the results~\cite{Baker+Steinke2}
of a semiclassical
expansion of the effective string theory \eqnlessref{Wilson Nambu Goto},
and use it to calculate Regge trajectories of mesons.

\section{Regge Trajectories of Mesons}

Consider an equal mass quark--antiquark pair separated by a
distance $R$ and rotating with angular velocity $\omega$.
The velocity of the quarks is $v = \omega R/2$.
For massless quarks, the ends of the string move with
the velocity of light, and singularities appear in the
expression \eqnlessref{Wilson Nambu Goto} for $W[\Gamma]$.
To regulate these singularities, we retain the quarks mass $m$ as a cutoff, and
take the limit $m\to 0$ at the end when evaluating physical
quantities.
\suppressfloats[t]
\begin {figure}[Ht]
    \begin{center}
	\null \hfill \epsfbox{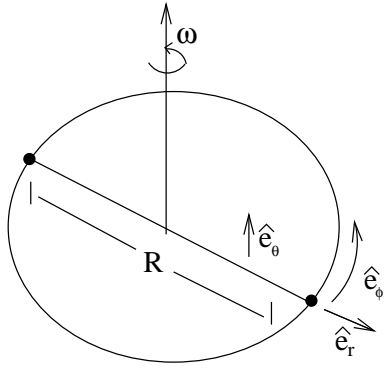} \hfill \null
    \end{center}
    \caption{The string coordinate system}
    \label{rot string fig}
\end {figure}
The effective Lagrangian for the rotating quark--antiquark
pair is given by
\begin{equation}
L(R, \omega) = -2m\sqrt{1-v^2} + L^{{\rm string}}(R, \omega) \,,
\label{total L}
\end{equation}
where
\begin{equation}
L^{{\rm string}} = \frac{-i}{T} \ln W[\Gamma] \,,
\label{string L}
\end{equation}
and where $T$ is the elapsed time.
We denote the straight string by $\bar x^\mu(r,t)$,
where the coordinate $r$ runs from $-R/2$ to $R/2$
(see Fig. \ref{rot string fig}). The semiclassical
expansion of $W[\Gamma]$ about $\bar x^\mu$
gives
\begin{equation}
L^{{\rm string}} = L^{{\rm string}}_{{\rm cl}} + L_{{\rm fluc}} \,,
\label{string cl L}
\end{equation}
where
\begin{equation}
L^{{\rm string}}_{{\rm cl}} = - \frac{\sigma}{T} \int d^2\xi \sqrt{- \bar g}
= -\sigma \int_{-R/2}^{R/2} dr \sqrt{1-r^2\omega^2} \,,
\label{L tension}
\end{equation}
and $L_{{\rm fluc}}$ is the contribution of the long wavelength
transverse vibrations of the rotating string, which is
analogous to the expression~\cite{Luscher1,Luscher2} of L\"uscher for
static strings.
We write \eqnlessref{total L} and \eqnlessref{string cl L} in the form
\begin{equation}
L(R,\omega) = L_{{\rm class}} + L_{{\rm fluc}} \,,
\end{equation}
with
\begin{equation}
L_{{\rm class}} = -2m\sqrt{1-v^2}
- \sigma \int_{-R/2}^{R/2} dr \sqrt{1-r^2\omega^2} \,.
\end{equation}

The expression for $L_{{\rm fluc}}$ obtained from
\eqnlessref{Wilson Nambu Goto} contains terms which are quadratically,
linearly, and logarithmically divergent in the cutoff $M = 1/a$.
The quadratic term is a renormalization of the string tension,
the linear term is a renormalization of the quark mass, and the
logarithmically divergent term is proportional to the integral
of the scalar curvature over the whole worldsheet.~\cite{Luscher1}
After absorbing the quadratic and linear terms into a renormalization
of $\sigma$ and $m$, we obtained,~\cite{Baker+Steinke2}
\begin{equation}
L_{{\rm fluc}} = \frac{\pi}{12 R_p} - \frac{2\omega v \gamma}{\pi}
\left[ \ln\left(\frac{MR}{2(\gamma^2-1)}\right) + 1 \right]
+ \frac{\omega}{2} + \omega f(v) \,,
\label{L fluc}
\end{equation}
where $\gamma = \gamma(v) = 1/\sqrt{1-v^2}$, $R_p$ is the
``proper length'' of the string,
\begin{equation}
R_p = \frac{2\arcsin v}{\omega} \,,
\end{equation}
and $f(v)$ is a function of the quark velocity which approaches zero
as $v \to 1$.

In the limit $\omega \to 0$, $L_{{\rm fluc}}$ \eqnlessref{L fluc} reduces to the
result of L\"uscher~\cite{Luscher2} for the corrections to the
static quark--antiquark potential due to string fluctuations,
\begin{equation}
V_{\hbox{\scriptsize L\"uscher}} = -L_{{\rm fluc}}(R, \omega = 0)
= - \frac{\pi}{12 R} \,.
\label{Luscher term}
\end{equation}
In this situation, the vortex is flat and $L_{{\rm fluc}}$ is divergence free,
since the divergent term  in \eqnlessref{L fluc} is proportional to $\omega$.
For $\omega \ne 0$, the vortex sheet is not flat, and there is
an ultraviolet logarithmic divergence in $L_{{\rm fluc}}$.
In this case we need to evaluate the classical equations of
motion to determine the relationship between $R$, $\gamma$, and $\omega$.

The classical equation of motion,
\begin{equation}
\frac{\partial L_{{\rm class}}}{\partial R} = 0 \,,
\end{equation}
gives the relation
\begin{equation}
\sigma \frac{R}{2} = m(\gamma^2 - 1) \,,
\label{boundary condition}
\end{equation}
which determines $R$ as a function of $\omega$,
\begin{equation}
R(\omega) = \frac{2}{\omega} \left( \sqrt{\left(\frac{m\omega}{2\sigma}\right)^2
+ 1} - \frac{m\omega}{2\sigma} \right) \,.
\label{R of omega}
\end{equation}
For $m=0$, \eqnlessref{R of omega} becomes $R = 2/\omega$, and $v = 1$.

Using \eqnlessref{boundary condition} to eliminate $R$ in
\eqnlessref{L fluc} gives
\begin{equation}
L_{{\rm fluc}} = \frac{\pi}{12 R_p} - \frac{2\omega v \gamma}{\pi}
\left[ \ln\left(\frac{Mm}{\sigma}\right) + 1 \right] + \frac{\omega}{2}
+ \omega f(v) \,.
\label{L fluc bc}
\end{equation}
The second term in $L_{{\rm fluc}}$, which contains the logarithmic
divergence, is equal to $\omega v \gamma$ multiplied by
a constant which is independent of $\omega$. The quantity 
$\omega v \gamma$ is equal to the geodesic
curvature~\cite{Alvarez:1983} (which measures 
the curvature of the quark
worldline in the plane of the string worldsheet) 
when evaluated for a straight, rotating string.
The logarithmic divergence in
\eqnlessref{L fluc bc} is then a renormalization of
the geodesic curvature, so we remove it from $L_{{\rm fluc}}$.
As the quark mass goes to zero the geodesic curvature $\omega v\gamma$
becomes singular. Therefore, in the massless quark limit the
renormalized coefficient of the geodesic curvature must vanish,
so it is not included in the classical Lagrangian.
Removing the term in \eqnlessref{L fluc bc} proportional to
$\omega v \gamma$ and setting
the quark mass $m$ to zero gives
\begin{equation}
L_{{\rm fluc}} = \frac{\omega}{12} + \frac{\omega}{2} = \frac{7}{12} \omega \,.
\label{L fluc massless}
\end{equation}

The result \eqnlessref{L fluc massless} for $L_{{\rm fluc}}$ is
readily generalized to $D$ dimensional spacetime, for which there
are $D-2$ transverse oscillations of the string.
The $\omega/12$ term in \eqnlessref{L fluc massless} arises from
the generalized L\"uscher term for rotating quarks
(the first term in \eqnref{L fluc bc}). Its dependence on $D$
is the same as that of the L\"uscher term for static quarks,
so that $1/12$ is replaced by $(D-2)/24$.

The $\omega/2$ term arises from the corrections to the
zero point energy of string fluctuations due to the
extrinsic curvature of the string.
This term is the sum of a contribution
from the zero point motion of the string
in the $\hat e_\phi$ direction, in the string's
plane of rotation, and a contribution from the
zero point motion of the string in the
$\hat e_\theta$ direction, perpendicular
to the plane of rotation (see Fig.~\ref{rot string fig}).
In the limit of massless quarks the contribution of
fluctuations in the $\hat e_\theta$ direction
is zero. Going from four to $D$ dimensions will
allow the string to fluctuate in additional directions
perpendicular to the plane of rotation, but,
like the fluctuations in the $\hat e_\theta$ direction,
these additional fluctuations will not affect
the contribution of the curvature of the string
to its energy.
Therefore, for a rotating string with massless quarks on its ends
in $D$ dimensional spacetime we obtain
\begin{equation}
L_{{\rm fluc}} = \frac{D-2}{24} \omega + \frac{\omega}{2} \,.
\label{L fluc general D}
\end{equation}

Comparing \eqnlessref{L fluc massless} with the classical value of the
Lagrangian,
\begin{equation}
L_{{\rm class}} = -\sigma \int_{-1/\omega}^{1/\omega} dr \sqrt{1-r^2\omega^2}
= -\frac{\pi\sigma}{2\omega} \,,
\end{equation}
gives
\begin{equation}
\frac{L_{{\rm fluc}}}{L_{{\rm class}}} = -\frac{7}{12J} \,,
\end{equation}
where
\begin{equation}
J = \frac{\partial L_{{\rm class}}}{\partial\omega}
= \frac{\pi}{2} \frac{\sigma}{\omega^2} \,,
\label{J of omega}
\end{equation}
is the classical angular momentum of the rotating string.

We consider $J \gg 1$, so that $L_{{\rm fluc}}$ can be treated
as a perturbation. The energy $E(\omega)$ of the rotating string
is then
\begin{equation}
E(\omega) = E_{{\rm class}}(\omega) - L_{{\rm fluc}}(\omega)
= E_{{\rm class}}(\omega) - \frac{7}{12} \omega \,,
\label{E of omega}
\end{equation}
where
\begin{equation}
E_{{\rm class}}(\omega) = \omega J - L_{{\rm class}}
= \frac{\pi\sigma}{\omega} \,,
\end{equation}
is the classical value of the energy.
Taking the square of \eqnlessref{E of omega} and retaining
terms of first order in $L_{{\rm fluc}}$ gives
\begin{equation}
E^2 = \frac{\pi^2\sigma^2}{\omega^2} - \frac{7}{12} (2\pi\sigma)
= 2\pi\sigma \left( J - \frac{7}{12} \right) \,,
\end{equation}
or
\begin{equation}
J = \frac{E^2}{2\pi\sigma} + \frac{7}{12} \,.
\label{shifted Regge}
\end{equation}
The attractive contribution $-7\omega/12$ of the string fluctuations to
the energy $E(\omega)$ has shifted the intercept of the
classical Regge trajectory $J = E^2/2\pi\sigma$ by $7/12$.
The generalization of \eqnlessref{shifted Regge} to $D$
dimensions, obtained from \eqnlessref{L fluc general D}, is
\begin{equation}
J = \frac{E^2}{2\pi\sigma} + \frac{D-2}{24} + \frac{1}{2} \,.
\label{general Regge}
\end{equation}

\section{Quantization of Angular Momentum}

The boundary $\Gamma$ of the fluctuating vortex is determined
by the frequency $\omega$ of the rotating quarks and the elapsed time $T$,
both of which are fixed by the classical trajectories of the quarks.
The frequency $\omega$ and the angular momentum $J$ are classical
variables which can take on any positive values related by
\eqnlessref{J of omega}. The quantum fluctuations of the ends of
the string are not included in $L_{{\rm fluc}}$.

To take these fluctuations into account we extend the
functional integral \eqnlessref{Wilson Nambu Goto} to include a path
integral over the coordinates of the ends of the string.
We then use the methods of Dashen, Hasslacher, and Neveu~\cite{DHN}
to carry out a semiclassical calculation of this
path integral around periodic classical solutions.
As a result, we obtain the WKB quantization condition
for angular momentum,
\begin{equation}
J = l + \frac{1}{2} \,, \kern 1in l = 0, 1, 2, ... \,.
\label{J of l}
\end{equation}
Note that the semiclassical shift
of the Regge trajectory \eqnlessref{general Regge} is the sum of two contributions,
one coming from the generalized L\"uscher term,
and one from the curvature of the string worldsheet.
The $1/2$ in \eqnlessref{J of l} cancels the
curvature contribution, and gives the Regge
trajectory
\begin{equation}
l = \frac{E^2}{2\pi\sigma} + \frac{D-2}{24}
+ O\left(\frac{\sigma}{E^2}\right) \,.
\label{quantized Regge D}
\end{equation}
The constant shift of $(D-2)/24$ to the angular momentum of mesons
on the leading Regge trajectory comes entirely from the relativistic
L\"uscher term $((D-2)/24)\omega$ in \eqnlessref{L fluc general D}.
In four dimensions, \eqnlessref{quantized Regge D} becomes
\begin{equation}
l = \frac{E^2}{2\pi\sigma} + \frac{1}{12}
+ O\left(\frac{\sigma}{E^2}\right) \,.
\label{quantized Regge}
\end{equation}

The expansion parameter in \eqnlessref{quantized Regge D} is
$(D-2)/{24l}$, or $1/{12l}$ for $D=4$. This parameter is already
small for $l=1$, so \eqnlessref{quantized Regge} should be applicable
to mesons with angular momenta $l \ge 1$. Furthermore, the small size,
$1/12$, of the first semiclassical correction could provide 
an explanation why the classical Regge formula works so well. 

\section{Regge Trajectories of Hybrid Mesons}

The hybrid mesons are excited states of the rotating string. The
string has two normal modes of frequency $k\omega$ for
each integer $k$ greater than zero. These two modes are
vibrations in the two directions $\hat e_\theta$
and $\hat e_\phi$ transverse to the classical string.
The direction $\hat e_\phi$ is in the plane of
the orbit, and the direction $\hat e_\theta$ is
perpendicular to the orbit (see Fig.~\ref{rot string fig}).
The ground state energy $E(\omega)$ includes the
zero point energy of these modes. The energies
$E_n(\omega)$ of the excited states are obtained by
adding $n\omega$ to the ground state energy \eqnlessref{E of omega},
\begin{equation}
E_n(\omega) = E_{cl}(\omega) - \frac{7}{12} \omega + n\omega \,.
\label{nth energy}
\end{equation}
There are two $n=1$ levels, each corresponding to
a single excitation of one of the $k=1$ normal modes.
There are five $n=2$ levels. Two are single excitations
of $k=2$ modes, two are double excitations of $k=1$ modes,
and one is a simultaneous single excitation of both $k=1$ modes.
For some of the smaller values of $n$, the
degeneracy of the hybrid mesons is
\vspace{1cm}
\begin{center}
	\setlength{\tabcolsep}{.1in}
	\begin{tabular}{|*{7}{l|}}
		\hline
		n & 0 (ground state) & 1 & 2 & 3 & 4 & 5 \\
		\hline
		degeneracy & 1 & 2 & 5 & 10 & 20 & 36 \\
		\hline
	\end{tabular}
\end{center}
\vspace{1cm}

The Regge trajectories of hybrid mesons can be obtained
from the energies \eqnlessref{nth energy} in the same
way that the lowest trajectory \eqnlessref{quantized Regge}
was obtained from the ground state energy \eqnlessref{E of omega}.
The hybrid trajectories are then
\begin{equation}
l = \frac{E^2}{2\pi\sigma} + \frac{1}{12} - n \,,
\kern 1 in n = 0, 1, 2, ... \,\,.
\label{excited trajectories}
\end{equation}
Higher order corrections to this semiclassical formula
are suppressed by a factor of $l^{-1}$, so corrections
to \eqnlessref{excited trajectories} will be of order
$n/l$. The trajectories \eqnlessref{excited trajectories}, for $n>0$,
are the daughters of the leading trajectory \eqnlessref{quantized Regge}.
They determine the spectrum of hybrid mesons of angular momentum
$l > n$, where \eqnlessref{excited trajectories} is applicable.
The degeneracy of a level on the $n$th trajectory is
given in the table.

In the usual flux tube model for hybrid
mesons,~\cite{Isgur+Paton,Allen+Olsson+Veseli:1998}
a heavy quark--antiquark pair separated by a distance $R$ are
held fixed, and the energy levels $E_n(R)$ of the quanta of
the vibrational modes of the flux tube with fixed ends
are used to determine excited potentials $V_n(R) \equiv -E_n(R)$.
Using these excited potentials in the Schr\"odinger equation
for a massive quark--antiquark pair then gives the spectrum
of hybrid mesons in the usual flux tube model.
It would be interesting to compare these energies to
those predicted by the semiclassical Regge trajectories 
\eqnlessref{excited trajectories}.

\section{Comparison with Bosonic String Theory}

We use the formula \eqnlessref{excited trajectories} to
make an illustrative comparison with classical bosonic string
theory.~\cite{GSW}
The generalization of \eqnlessref{excited trajectories}
to $D$ dimensional spacetime is
\begin{equation}
l = \frac{E^2}{2\pi\sigma} + \frac{D-2}{24} - n
+ O\left(\frac{\sigma}{E^2}\right) \,.
\label{D dim trajectories}
\end{equation}
In 26 dimensions, \eqnlessref{D dim trajectories}
yields the spectrum
\begin{equation}
E^2 = 2\pi\sigma\left(N - 1 + O(l^{-1}) \right) \,,
\label{26 dim spectrum}
\end{equation}
where $N = l + n$. The spectrum of energies \eqnlessref{26 dim spectrum}
coincides with
the spectrum of open strings in bosonic string theory in its
critical dimension $D=26$. However, in our approach, \eqnlessref{26 dim spectrum}
is valid only in the leading semiclassical approximation, so that it cannot
be used for $l=0$, where it would yield the scalar tachyon of bosonic
string theory.

\section{Comparison with the Classical Regge Formula}

\begin {figure}[Ht]
    \begin{center}
	\begin{tabular}{rclc}
	    \vbox{\hbox{$l+1$ \hskip 0in \null} \vskip 1.2 in} &
	    \epsfxsize = 3.4 in
	    \epsfbox{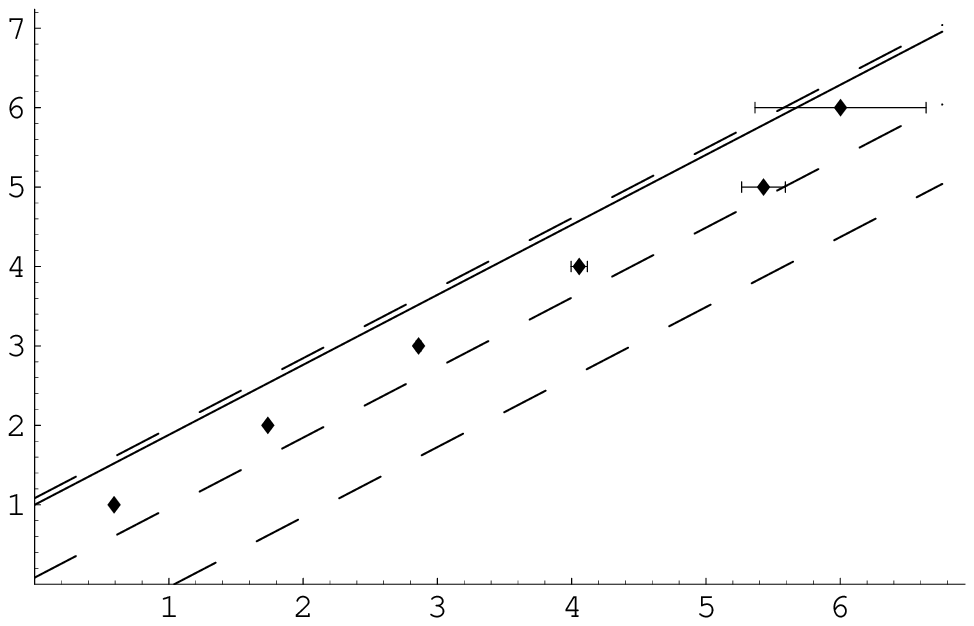} &
	    \vbox{\hbox{$n$} \vskip 0.22 in
		\hbox{0} \vskip 0.2 in \hbox{1} \vskip 0.2 in
		\hbox{2} \vskip 1.55 in} &
	    \vbox{\hbox{degeneracy} \vskip 0.2 in
		\hbox{1} \vskip 0.2 in \hbox{2} \vskip 0.2 in
		\hbox{5} \vskip 1.55 in} \\
	    &
	    \hbox{$E^2$ in GeV$^2$} &
	    &
	    \\
	\end{tabular}
    \end{center}
    \caption{Dotted lines: Semiclassical Regge trajectories 
\eqnlessref{excited trajectories} with $n=0,1,2$.
Solid line: Classical Regge trajectory.
Points: Measured masses of mesons on the $\rho$ - $a_2$ trajectory.}
    \label{Regge plot}
\end {figure}

In Fig.~\ref{Regge plot}, we plot the leading Regge trajectory and the
first two daughters (\eqnref{excited trajectories} with $n = 0,1,2$)
using the value $\sigma = (\hbox{425 MeV})^2$. For comparison,
we also plot the classical formula $J = E^2/2\pi\sigma$.
The plotted points are those on the $\rho$ - $a_2$ trajectory.
We have added one to the orbital angular momentum $l$ to
account for the spin of the quarks ($J = l + s = l + 1$). 
The plotted points lie  to the right of
the leading trajectory, so the predicted masses are 
too low.

The Regge trajectories in Fig.~\ref{Regge plot} were derived using
massless quarks. The longitudinal mode (the motion of the ends 
of the string along the vector $\hat e_r$ shown in 
Fig.~\ref{rot string fig}) did not contribute to $L_{fluc}$.
(For massless quarks the longitudinal mode is unphysical.)
For massive quarks this mode will contribute to the energy
of mesons and should increase their masses, so
mesons containing massive quarks should have larger masses
than our predictions. In fact, quantum mechanical treatments
~\cite{LaCourse+Olsson,Dubin+Kaidalov+Simonov}
of the radial motion of a straight string give meson masses
which are too large. Taking into 
account the effect of string
fluctuations would lower these masses.

\section{Summary and Conclusions}

\begin{enumerate}
    \item We have shown how a field theory containing classical vortex
	solutions can be expressed as an effective string theory
	of long distance QCD. The effective string theory
	\eqnlessref{Wilson Nambu Goto} is a two dimensional field
	theory describing the two transverse
	oscillations of the string. In $D$ dimensional spacetime,
	there are $D-2$ transverse oscillations.
	The action of the effective string theory is the Nambu--Goto
	action, and short distance fluctuations of the string are
	cutoff at the flux tube radius.
    \item The semiclassical expansion of the effective string
	theory about a classical rotating string solution gives
the contribution of string fluctuations to meson
	Regge trajectories. The semiclassical correction to the
leading trajectory adds the constant $1/12$ to the classical 
Regge formula. This is a basic new result of our work.
    \item The excited states of the rotating string give rise to
	daughter Regge trajectories \eqnlessref{excited trajectories}
	determining the spectrum of hybrid mesons 
composed of zero mass quarks. These calculations are
carried out in the semiclassical approximation where the
effective string theory is applicable to QCD.
\end{enumerate}

\section*{Acknowledgements}

We would like to thank N. Brambilla and P. Bicudo  for many very helpful
discussions, and for pointing out the connection of our
approach to other work on hybrid mesons.



\end{document}